\documentclass[10pt,journal,twocolumn,twoside]{IEEEtran}
\usepackage{amsthm,amsmath,amsfonts,amsbsy}
\usepackage{mathtools}
\usepackage{bm}
\usepackage{cite}
\usepackage[caption=false,font=footnotesize]{subfig}
\usepackage[nolist]{acronym}
\usepackage{balance}

\mathtoolsset{showonlyrefs}

\newtheorem{example}{Example}

\newcommand{\PUPE}{\mathsf{PUPE}}
\def\PP{\mathbb{P}}
\def\triangleq{:=}

\IEEEoverridecommandlockouts

\begin{document}

%%%%%%%%%% title section %%%%%%%%%%%%%%%%%%%%%%%%%%%%%%%%%%%%%%%%%%%%%%%%%%%%%%%%%%%%%
%%%%%%%%%%%%%%%%%%%%%%%%%%%%%%%%%%%%%%%%%%%%%%%%%%%%%%%%%%%%%%%%%%%%%%%%%%%%%%%%%%%%%%
%%%%%%%%%%%%%%%%%%%%%%%%%%%%%%%%%%%%%%%%%%%%%%%%%%%%%%%%%%%%%%%%%%%%%%%%%%%%%%%%%%%%%%

\title{Unsourced Multiple Access: A Coding Paradigm for Massive Random Access}
\author{Gianluigi Liva, \IEEEmembership{Senior Member, IEEE}, and Yury Polyanskiy, \IEEEmembership{Fellow, IEEE}
	\thanks{
		G. Liva is with the Institute of Communications and Navigation, German Aerospace Center (DLR), Wessling, Germany (email: gianluigi.liva@dlr.de).
	}
	\thanks{
		Y. Polyanskiy is with the Department of Electrical Engineering and Computer Science, Massachusetts Institute of Technology, Cambridge, MA, 02139 USA (email: yp@mit.edu).   
	}
	\thanks{
		G. L. acknowledges the financial support by the Federal Ministry of Education and
		Research of Germany in the framework of the ``Souver\"an. Digital. Vernetzt.'' Joint
		project 6G-RIC, project identification number: 16KISK022. The work of Y. P. was
		supported by the National Science Foundation under Grant No CCF-2131115.
	}
}

\maketitle
\thispagestyle{empty}

\markboth
{Accepted for Publication, Proceedings of the IEEE}
{G. Liva, Y. Polyanskiy: Unsourced Multiple Access: A Coding Paradigm for Massive Random Access}

\maketitle

\IEEEoverridecommandlockouts

\begin{abstract}
This paper is a tutorial introduction to the field of unsourced multiple access (UMAC) protocols. We first provide a historical survey of the evolution of random access protocols, focusing specifically on the case in which uncoordinated users share a wireless broadcasting medium. Next, we highlight the change of perspective originated by the UMAC model, in which the physical and medium access layer's protocols cooperate, thus reframing random access as a novel coding-theoretic problem. By now, a large variety of UMAC protocols (codes) emerged, necessitating a certain classification that we indeed propose here. Although some random access schemes require a radical change of the physical layer, others can be implemented with minimal changes to existing industry standards. As an example,  we discuss a simple modification to the 5GNR Release 16 random access channel that builds on the UMAC theory and that dramatically improves energy efficiency for systems with even moderate number of simultaneous users (e.g., $5-10$ dB gain for $10-50$ users), and also enables handling of high number of users, something completely out of reach of the state-of-the-art.
\end{abstract}


%%%%%%%%%% introduction %%%%%%%%%%%%%%%%%%%%%%%%%%%%%%%%%%%%%%%%%%%%%%%%%%%%%%%%%%%%%%
%%%%%%%%%%%%%%%%%%%%%%%%%%%%%%%%%%%%%%%%%%%%%%%%%%%%%%%%%%%%%%%%%%%%%%%%%%%%%%%%%%%%%%
%%%%%%%%%%%%%%%%%%%%%%%%%%%%%%%%%%%%%%%%%%%%%%%%%%%%%%%%%%%%%%%%%%%%%%%%%%%%%%%%%%%%%%

\section{Introduction}\label{sec:intro}

\IEEEPARstart{T}{he} breakneck development of wireless data networks witnessed during the past decades has been continuously driven by new applications. In its early implementations, wireless cellular networks targeted email and messaging services, which required moderate to low data rates. High resolution multimedia content on the Internet and two-way video streaming introduced the need for broadband connectivity. Recently, the rise of \ac{mMTC} and \ac{IoT} systems placed a new set of challenges for the design of next-generation wireless systems. These latter use cases entail drastically different features in terms of traffic profile and reliability requirements. As a consequence, new technical solutions that can address the peculiarities of \ac{mMTC} and \ac{IoT} systems have been the subject of intense research efforts in recent years. From a medium access point of view, the shift in perspective originating from these new applications is today well understood.  Focusing on traffic profiles only, \ac{mMTC} and \ac{IoT} systems often foresee large populations of terminals, which are active sporadically and unpredictably and that transmit only small datagrams. This is in stark contrast with the classical setting of broadband connectivity, where the terminal population is typically orders of magnitude smaller than the one of \ac{mMTC} and \ac{IoT} systems, and the data exchange between a user and the \ac{BS} involves the transmission of large amounts of data, allowing the use of efficient scheduling techniques to handle medium access. Originally, random access protocols (Aloha in~\cite{Abramson:ALOHA}) emerged as a technique to enable wireless access connectivity without centralized coordination between users. Although the development of random access techniques predates by several decades the development of cellular networks, they still form a crucial part of modern 5G stacks for the purpose of providing initial access and handling resource requests. Even if perfectly adequate for those use cases in the past, with \ac{mMTC}/\ac{IoT} systems random access shall become the main mechanism for transmitting data, thus placing a much stronger emphasis on the need for energy/spectrum efficient protocols and necessitating revamping of the old designs.

The new challenges placed by \ac{mMTC}/\ac{IoT} systems led to an information-theoretic treatment of \ac{MRA} in \cite{Polyanskiy2017}, gave rise to an actively developing field of \ac{UMAC}, and revived interest in the design of random access schemes. The paradigm shift manifested by the \ac{UMAC} is conceptually simple: Instead of relegating the details of random access to the medium access control layer, one should leverage additional side information obtained from the physical layer. In doing so, the random access problem can be formulated as a \emph{coding problem} \cite{Polyanskiy2017}. It is important to emphasize that \ac{UMAC} provides a foundation to construct powerful \ac{MRA} schemes for \emph{uncoordinated} uplink channels. This is different from a wide body of work on next generation multiple access (NGMA) protocols for \emph{coordinated} downlink / uplink transmissions, which can be addressed by orthogonal and/or  non-orthogonal multiple access (NOMA) schemes. The latter is outside of the scope of this paper as we are exclusively focused on the uncoordinated uplink.

\subsection{Objective and Main Contributions}

In this paper, we pursue three objectives, namely:
\begin{itemize}
	\item[A.] We aim at providing a tutorial introduction to the field of \ac{MRA} protocols. We provide a historical survey on the development of random access protocols, which culminates with the introduction of the information theoretic treatment of \ac{MRA} provided by the \ac{UMAC} framework. 
	\item[B.] We discuss recent progress in the field of \ac{MRA} protocols, with emphasis on schemes that embrace the \ac{UMAC} perspective. We provide a classification of some of the most promising \ac{UMAC} architectures of recent introduction, highlighting their distinctive features.
	\item[C.] Finally, we will outline how these theoretical developments may influence the design of future (3GPP) wireless cellular systems. In particular, we show how simple modifications \cite{whitepaper2024} of the two-step random access protocol --- a grant-free random access protocol that has recently been introduced in the \ac{5GNR} standard \cite{5GNR16} --- can dramatically improve its efficiency, paving the way for the support of \ac{MRA} in future versions of the standard.\footnote{We mention that although the paper is written with an eye towards cellular networks (3GPP), almost everything we discuss carries over without change to the \textit{low-power wide-area networks (LP-WANs)}, such as LoRaWAN~\cite{Loraspec}, mioty~\cite{mioty} and Zigbee~\cite{Zigbee}.}
\end{itemize}

\subsection{Outline}

The contribution is structured as follows. Section \ref{sec:history} reviews the historical progress in the theory and practice of random access protocols, discussing the blurring of the separation between the medium access control layer and the physical layer. Section \ref{sec:UMAC} discusses the \ac{UMAC} framework, highlighting the implications of merging the medium access control layer and the physical layer from a coding theory viewpoint. Emerging \ac{UMAC} coding architectures are presented in \ref{sec:architectures}. The grant-free random access capabilities included in the \ac{5GNR} standard are illustrated in Section \ref{sec:5GNR}. Their limitations are discussed and possible directions for future developments are identified. Conclusions follow in Section \ref{sec:conc}.

\begin{acronym}
	\acro{5GNR}{$5$G New Radio}
	\acro{AMP}{approximate message passing}
	\acro{AWGN}{additive white Gaussian noise}
	\acro{BAC}{binary adder channel}
	\acro{BP}{belief propagation}
	\acro{BS}{base station}
	\acro{CDMA}{code division multiple access}
	\acro{CRDSA}{contention resolution diversity slotted Aloha}
	\acro{CS}{compressed sensing}
	\acro{CCS}{coded compressed sensing}
	\acro{CSA}{coded slotted Aloha}
	\acro{CSMA}{carrier sense multiple access}
	\acro{DVB}{Digital Video Broadcasting}
	\acro{E-SSA}{enhanced spread spectrum Aloha}
	\acro{FDMA}{frequency division multiple access}
	\acro{IDMA}{interleave division multiple access}
	\acro{IRSA}{irregular repetition slotted Aloha}
	\acro{IoT}{Internet of Things}
	\acro{LDPC}{low-density parity-check}
	\acro{LTE}{Long Term Evolution}
	\acro{MAC}{multiple access}
	\acro{MIMO}{multiple-input multiple-output}
	\acro{MMSE}{minimum mean squared error}
	\acro{MRA}{massive random access}
	\acro{mMTC}{massive machine-type communication}
	\acro{MPR}{multipacket reception}
	\acro{MUD}{multiuser detection}
	\acro{OMP}{orthogonal matching pursuit}
	\acro{PRACH}{physical random access channel}
	\acro{PUPE}{per-user probability of error}
	\acro{PUSCH}{\emph{physical uplink shared channel}}
	\acro{QPSK}{quadrature phase shift keying}
	\acro{RCS}{Return Channel via Satellite}
	\acro{SB-IDMA}{sparse block interleave division multiple access}
	\acro{SCL}{successive cancellation list}
	\acro{SNR}{signal-to-noise ratio}
	\acro{SPARC}{sparse regression code}
	\acro{SIC}{successive interference cancellation}
	\acro{TDMA}{time division multiple access}
	\acro{TIN}{treat interference as noise}
	\acro{UMAC}{unsourced multiple access}
\end{acronym}

\newcommand{\acrot}[2]{#1 & #2 \\}
\begin{table}[h]
	\centering
	\caption{List of Acronyms}
	\begin{tabular}{ll}
		\hline\hline\\[-2.5mm]
		\acrot{5GNR}{$5$G New Radio}
		\acrot{AMP}{approximate message passing}
		\acrot{AWGN}{additive white Gaussian noise}
		\acrot{BAC}{binary adder channel}
		\acrot{BP}{belief propagation}
		\acrot{BS}{base station}
		\acrot{CDMA}{code division multiple access}
		\acrot{CRDSA}{contention resolution diversity slotted Aloha}
		\acrot{CS}{compressed sensing}
		\acrot{CCS}{coded compressed sensing}
		\acrot{CSA}{coded slotted Aloha}
		\acrot{CSMA}{carrier sense multiple access}
		%\acrot{DVB}{Digital Video Broadcasting}
		\acrot{E-SSA}{enhanced spread spectrum Aloha}
		\acrot{FDMA}{frequency division multiple access}
		\acrot{IDMA}{interleave division multiple access}
		\acrot{IRSA}{irregular repetition slotted Aloha}
		\acrot{IoT}{Internet of Things}
		\acrot{LDPC}{low-density parity-check}
		\acrot{LTE}{Long Term Evolution}
		\acrot{MAC}{multiple access}
		\acrot{MIMO}{multiple-input multiple-output}
		\acrot{MMSE}{minimum mean squared error}
		\acrot{MRA}{massive random access}
		\acrot{mMTC}{massive machine-type communication}
		\acrot{MPR}{multipacket reception}
		\acrot{MUD}{multiuser detection}
		\acrot{OMP}{orthogonal matching pursuit}
		\acrot{PRACH}{physical random access channel}
		\acrot{PUPE}{per-user probability of error}
		\acrot{PUSCH}{physical uplink shared channel}
		\acrot{QPSK}{quadrature phase shift keying}
		\acrot{RCS}{Return Channel via Satellite}
		\acrot{SB-IDMA}{sparse block interleave division multiple access}
		\acrot{SCL}{successive cancellation list}
		\acrot{SNR}{signal-to-noise ratio}
		\acrot{SPARC}{sparse regression code}
		\acrot{SIC}{successive interference cancellation}
		\acrot{TDMA}{time division multiple access}
		\acrot{TIN}{treat interference as noise}
		\acrot{UMAC}{unsourced multiple access}
		\acrot{UT}{user terminal}
		\hline \hline
	\end{tabular}
	\label{tab:acronyms}
\end{table}

%%%%%%%%%% hystory %%%%%%%%%%%%%%%%%%%%%%%%%%%%%%%%%%%%%%%%%%%%%%%%%%%%%%%%%%%%%%%%%%%
%%%%%%%%%%%%%%%%%%%%%%%%%%%%%%%%%%%%%%%%%%%%%%%%%%%%%%%%%%%%%%%%%%%%%%%%%%%%%%%%%%%%%%
%%%%%%%%%%%%%%%%%%%%%%%%%%%%%%%%%%%%%%%%%%%%%%%%%%%%%%%%%%%%%%%%%%%%%%%%%%%%%%%%%%%%%%

\section{From Aloha to Codes for Unsourced Multiple Access}\label{sec:history}

We start with a brief perspective on the development of random access protocols. Our emphasis is on schemes developed in the context of wireless (including satellite) communications with an aim towards large-scale \ac{mMTC}/\ac{IoT} systems. Consequently, we will mostly ignore protocols that rely on carrier sensing, and protocols making intense use of feedback channels (such as splitting / contention tree algorithms).

\begin{figure*}
	\centering
	\includegraphics[width = 0.95\textwidth]{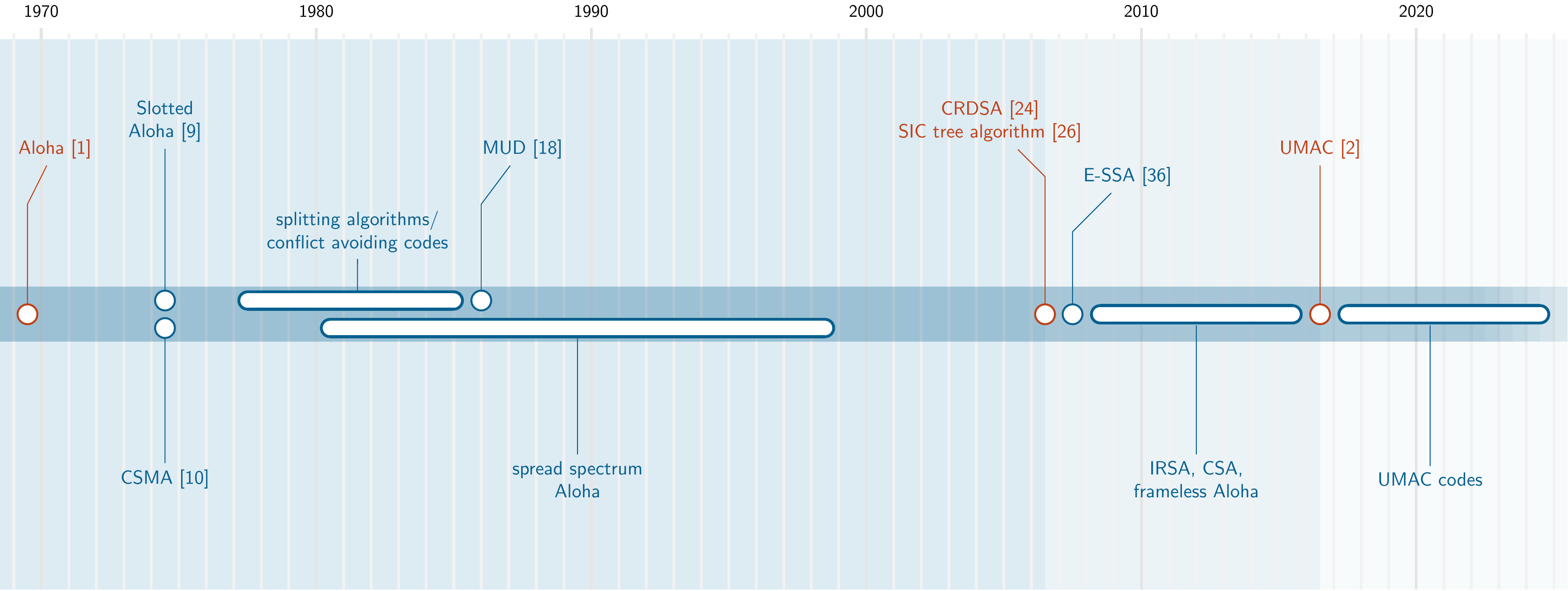}
	\caption{Timeline of the development of random access protocols.}
	\label{fig:timeline}
\end{figure*}

This section somewhat artificially divides the evolution of random access into three periods (see also Figure \ref{fig:timeline}). The first period (1970-2007) is dominated by Aloha-like schemes. The second period (2007-2017) builds on the introduction of \ac{MUD} techniques to improve the performance of random access protocols. The third period (starting in 2017, and still in progress) sees a paradigm change with the introduction of the \ac{UMAC} model, in which random access is viewed from a coding-theoretic perspective.\footnote{The choice of the dates delimiting the three periods is, of course, subjective. As a criterion, we decided to adopt the publication date of landmark papers that signal a change of perspective on the random access problem.}

\subsection{First Period (1970--2007): Aloha and Collision Models}\label{subsec:FirstPeriod} 

Initially, random access was understood as a layer-2 task, more specifically as part of the medium access control sublayer of the data link layer \cite{Bertsekas1992}. The introduction of Aloha \cite{Abramson:ALOHA} and of its slotted version \cite{Roberts1975} sets the ground for the development of sophisticated variations on the theme. These include \ac{CSMA} \cite{kleinrock1975packet}, splitting/contention-tree algorithms \cite{tsybakov1978free,Capetanakis} (see also [Chapter 4.3]\cite{Bertsekas1992}), and conflict avoiding codes \cite{MasseyCCwF,bassalygo1983restricted,tsybakov2002some}. Here, mutual interference among users is treated as destructive (hence, the notion of \emph{collisions}), and protocols aim at avoiding collisions (as for \ac{CSMA}), at resolving collision events via retransmissions (as in splitting/contention-tree algorithms), or controlling the number of collisions (as for conflict avoiding codes). The possibility of decoding in the presence of interference--- yielding the so-called \ac{MPR} capability \cite{Ghez1988}---is mostly considered in terms of \emph{capture effect}, i.e., in contexts where user transmissions arrive at the receiver antenna with a large difference in power \cite{Roberts1975,Abramson77:PacketBroadcasting,Ghez1988}. In this case the \ac{MPR} capability does not stem from a receiver design based  on \ac{MUD} techniques \cite{verdu1986minimum,Verdu1998}, but rather from propagation conditions in the multiple access channel. A significant departure from the collision model is represented by random access protocols that rely on spread spectrum techniques \cite{Raychaudhuri81,Gallager85:perspective,Polydoros87,Abramson90:VSAT}, where \ac{MPR} is explicitly targeted by signal design. An important example of a random access protocol based on spread spectrum waveforms is the spread Aloha protocol \cite{Abramson90:VSAT}, which will emerge as a high performance random access scheme when coupled with \ac{MUD} during the second period (see the following subsection). Although the benefits of this class of protocols were immediately recognized, Aloha and \ac{CSMA} dominated as the main random access techniques in wireless and wired data networks.

\subsection{Second Period (2007--2017): Aloha-like protocols with Multi-User Detection}
The second period coincides with the adoption of \ac{MUD} to improve the performance of Aloha-based algorithms. In this context, a close interaction is established between the medium access control layer and the physical layer by modifying the Aloha protocol to facilitate the application of multiuser signal processing techniques. 
A first example is the \ac{CRDSA} protocol \cite{DeGaudenzi07:CRDSA}, introduced to provide efficient use of satellite return channels in machine-type applications, and currently in use by the \ac{DVB} \ac{RCS} standard \cite{ETSI09b}. \ac{CRDSA} relies on packet repetition and on \ac{SIC} to improve the performance of slotted Aloha, providing tangible gains, especially at low packet error rates. A similar principle was devised in \cite{Giannakis07:SICTA} in the context of contention tree algorithms. It was soon recognized that the performance of \ac{CRDSA} under \ac{SIC} can be analyzed by establishing an analogy with erasure decoding of \ac{LDPC} codes \cite{Gal63}, providing the means to optimize the protocol behavior \cite{Liva11:IRSA}. In \cite{Liva2015:CodedAloha} a general protocol based on the \ac{CRDSA} principle --- named \ac{CSA} --- was introduced. The \ac{CRDSA} scheme of \cite{DeGaudenzi07:CRDSA} and the \ac{IRSA} scheme of \cite{Liva11:IRSA} can be recognized as special instances of \ac{CSA}. In \cite{Narayanan2012} it was shown that a suitable design of the \ac{IRSA} protocol allows achieving a peak throughput of one packet per slot, in the limit of large \ac{MAC} frame lengths. Variations of \ac{CSA} include the adoption of a feedback-based frameless approach \cite{Stefanovic12:Frameless,Stefanovic13:RatelessAloha} and spatial coupling \cite{Sandgren2017}, as well as the elimination of the assumption of a slotted frame structure \cite{DeG2014,Clazzer2018}. 

In parallel to \ac{CSA} techniques, \ac{MUD} applied to spread Aloha represented a key development of this period. Similarly to the case of \ac{CSA}, spread Aloha was studied mainly in the context of satellite \ac{mMTC}/\ac{IoT} networks. The \ac{E-SSA} protocol \cite{del_rio_herrero_high_2008} improves the \ac{MPR} capability of spread Aloha by canceling the interference contribution of decoded packets. Due to its completely asynchronous operation, its outstanding performance and lean transmitter/receiver design, \ac{E-SSA} emerged as a high-efficiency random access solution for \ac{mMTC}/\ac{IoT} and interactive satellite networks \cite{etsi_smim,Scalise13}.

\subsection{Third Period (2017--): Random Access as a Coding Problem}

The development of random access protocols during the first two periods has been largely based on a medium access control layer perspective. Consequently, development focused on packet-level metrics such as throughput, goodput and latency, whereas in the \ac{mMTC}/\ac{IoT} domain it was important to also address energy efficiency, thus requiring adequate modeling of the physical layer part. In addition, there was no model capable of capturing the fundamental aspects that differentiate random access from coordinated \ac{MAC}.

The \ac{UMAC} model \cite{Polyanskiy2017} resolves both of these issues and offers an information-theoretic ground to study random access schemes. This is achieved by recasting the problem into a coding-theoretic language. We will formally introduce \ac{UMAC} model in Section \ref{sec:UMAC}, but it is instructive to first consider the following example illustrating how random access can be seen as a ``coding'' problem in this framework.

\begin{example}[Slotted Aloha as \ac{UMAC} code] Let us consider transmission with framed slotted Aloha. The frame, consisting of $n$ complex channel uses, is divided into $L$ slots of $n_c$ complex channel uses each. According to the slotted Aloha protocol, an active user encodes its $k$-bits message into a word $\bm{w}$ of $n_c$ symbols via an $(n_c,k)$ block code $\mathcal{C}$, then it selects a slot to transmit the word $\bm{w}$. From a coding  point of view, we can describe the encoding performed by the user as the selection of a codeword with the form \begin{equation} \bm{x} = (\bm{x}_1, \bm{x}_2, \ldots, \bm{x}_L) \label{eq:SA} \end{equation} where $\bm{x}_\ell = \bm{w}$ if the user selected the $\ell$th slot, and $\bm{x}_\ell = \bm{0}$ (length-$n_c$ zero vector) otherwise. Hence, the \textit{codebook} realized by the slotted Aloha protocol is given by all $n$-tuples in the form \eqref{eq:SA} where one block is a word from $\mathcal{C}$ and all the other blocks are zero vectors, i.e., slotted Aloha can be seen as the \ac{UMAC} code \[ \mathcal{C}_{\emph{\textsf{SA}}}\! =\! \left\{(\bm{x}_1, \bm{x}_2, \ldots, \bm{x}_N)\! \in\! \mathbb{C}^n | \bm{x}_\ell\! \in\! \mathcal{C}, \bm{x}_j\! =\! \bm{0}, j\neq \ell, \ell\! \in \![L]\right\}. \] Each active user transmits a codeword from $\mathcal{C}_{\emph{\textsf{SA}}}$. The cardinality of the slotted Aloha codebook is therefore $|\mathcal{C}_{\emph{\textsf{SA}}}| = L|\mathcal{C}|$. The selected slot can either be random (as in the original slotted Aloha) or it could be chosen by computing a hash function of the payload data, thus providing extra parity checks for decoder. \end{example}

As the example shows, in the \ac{UMAC} framework, the ``code'' incorporates aspects of both the physical layer and the medium access control layer. We found that presenting this example is crucial for explaining the idea of \ac{UMAC} codes to network engineers, since they traditionally considered codes to operate only at the physical layer, and not think of the medium access control protocol part as ``code''. However, as we can see in the case of slotted Aloha, the selection of the slot used for the transmission can be thought of as rudimentary code. The inclusion of control layer aspects in \ac{UMAC} is an essential step in building a unifying theory of \ac{MRA} protocols, allowing a fair comparison of several \ac{MAC} strategies.

With this preview, we are ready to introduce the \ac{UMAC} model (Section \ref{sec:UMAC}) and discuss emerging practical approaches (Section \ref{sec:architectures}).

%%%%%%%%%% coding perspective %%%%%%%%%%%%%%%%%%%%%%%%%%%%%%%%%%%%%%%%%%%%%%%%%%%%%%%%
%%%%%%%%%%%%%%%%%%%%%%%%%%%%%%%%%%%%%%%%%%%%%%%%%%%%%%%%%%%%%%%%%%%%%%%%%%%%%%%%%%%%%%
%%%%%%%%%%%%%%%%%%%%%%%%%%%%%%%%%%%%%%%%%%%%%%%%%%%%%%%%%%%%%%%%%%%%%%%%%%%%%%%%%%%%%%

\section{Random access from an information-theoretic perspective}\label{sec:UMAC}

Let us start with describing the standard \ac{MAC} setting as studied in communication theory and information theory: \begin{itemize} \item \textit{Uplink:} A single \ac{BS} transmitting beacons and listening on a common broadcast channel for the uplink transmissions.
	
	\item \textit{Users:} Multiple users $K$ ($K$ at most a few hundred) are simultaneously using the uplink channel. The identities of the communicating users are known to the BS (due to prior control plane exchanges).
	
	\item \textit{Payload:} The users are sending either continuous streams (voice communication) or a large number of information bits in each session (data transfer).
	
	\item \textit{Multiple access:} Before the users are allowed to join the uplink communication, they have to announce their existence to the BS via a different channel --- known as \ac{PRACH} in the \ac{LTE} standard. Upon establishing their communication intent, the \ac{BS} instructs all currently active users on how they should share the channel access (in \ac{LTE} the \ac{BS} schedules resource blocks and configures time-offsets). The resulting allocations are distributed to the users as part of the beacon broadcast. \end{itemize} We note that in this setting the uplink multiple access is completely \textit{coordinated} by the BS. Because each user is sending a very large payload, the overhead that it spent on the resource acquisition phase and coordination are amortized.  From an information-theoretic point of view, thus, the main challenge in the setting above is that of designing $K$ (different!) channel codes, such that when $K$ random codewords (one codeword from each) are transmitted simultaneously on the uplink, the BS is able to recover each of them with a high probability of success. How to choose these $K$ codes is the subject of the information-theoretic \ac{MAC}, see~\cite[Section 15.3]{CT}. However, for the additive Gaussian noise channels, it is known that using a single standard (point-to-point) error-correcting code and simply allocating non-overlapping time-frequency resources to different users is optimal from the channel capacity point of view.

Next, let us describe the \ac{mMTC}/\ac{IoT} communication setting. Specifically, we will consider the following assumptions: \begin{itemize} \item \textit{Uplink:} A single base station (BS) transmitting beacons and listening on a common broadcast channel for the uplink transmissions.
	
	\item \textit{Users:} A very large (order a million) number of users (mMTC/IoT devices), of which majority are idle. When idle, users do not monitor the BS transmissions to conserve battery. Despite a large total number of users, only $K_a$ of them  have any data to send on the uplink ($K_a$ again is on the order of a few hundred).
	
	\item \textit{Payload:} When users have data to send, their messages are rather short ($100$s of bits). The message payload may simply be an identity (cryptographically signed) of the user, or the identity plus a short status update.
	
	\item \textit{Random access:} Users wake up from the idle state at random, without BS knowing who is awake at any given time. Since the duty cycle (the amount of time the user has to stay on before completing its radio access) directly contributes to battery depletion, the desire of each users is to initiate communication immediately after waking up. \end{itemize}

Comparing the two settings side by side clearly shows the salient feature of the latter: While the total number of communicating users may be roughly the same ($K\approx K_a$), the identities of communicating users in the second setting are unknown. Correspondingly, the communication needs to proceed in a completely uncoordinated way. How is it possible to achieve any reliable data transmission by multiple users without them coordinating in some fashion?

The simplest and most ubiquitous solution is the Aloha protocol: Each user, whenever it has data to send, simply transmits its message on the uplink (the \ac{CSMA} variation requires the user to first check that the uplink is idle and, if not, to retry at a random later time). We see that if transmissions of the users are very short and wake-up times are random, the chance of collisions is low. Let us try to estimate this chance.

Suppose that at the beginning of a frame (following a beacon) we have $K_a$ users ready to transmit their data.  We can slice our frame into $L$ nonoverlapping slots\footnote{Instead of this TDMA-like idea, we could also divide the channel according to some other orthogonal basis. For example, for random access in LTE (PRACH) the users choose from $L=64$ possible Zadoff-Chu sequences, which all overlap in time but otherwise are orthogonal.}. In accordance with the Aloha principle, each of $K_a$ users selects one of the $L$ slots at random and places its message there. In this case, let us fix one of the users and ask what is the probability that someone else selects its slot for communication: $$ P_{\mathrm{collision}} = 1-\left(1-{1\over L}\right)^{K_a-1}  \le {K_a -1 \over L}\,.$$ This calculation implies the following important conclusion: Unless we are able to decode the packets that collide (are transmitted in the same slot) there is an error floor $\approx {K_a-1\over L}$ for the probability of recovering a user's message. Since making $L$ very large is impractical, we are led to the natural conclusion that any random access scheme must use some \ac{MPR} capabilities.

Let us summarize our findings. For realistic values of $L$, the Aloha principle alone is not capable of producing an uncoordinated random access with low probability of error. All of the users who selected the same slot (or the same Zadoff-Chu preamble in LTE's PRACH) appear completely identical to the BS. That is, from the point of view of the BS, these users are all employing an identical transmission strategy, with the only difference that each of them is transmitting a different payload. How to produce such a transmission strategy is precisely the topic of \textit{UMAC coding theory} (with ``U'' standing for uncoordinated or unsourced).

We now describe more formally what a UMAC code is meant by. First, we define the two channel models that we use in the remainder of the paper. Second, we recall the definition of a point-to-point error-correcting code. Third, we define the UMAC code.

\textbf{Channel models.} We only consider single antenna channels. Recall that a $K$-user \ac{AWGN} channel with $K$ users and blocklength $n$ takes as input $K$ vectors $\bm{x}_1,\ldots,\bm{x}_K \in \mathbb{C}^n$ and outputs a random $\bm{Y} \in \mathbb{C}^n$ according to $$\bm{Y} = \sum_{i=1}^K \bm{x}_i + \bm{Z}, \qquad \bm{Z} \sim \mathcal{CN}(\bm{0},\sigma^2 \bm{I}_n)$$ where $\bm{Z}$ is a complex vector with $n$ i.i.d. complex normal entries of power $\sigma^2$ per entry. All channel inputs are subject to a power constraint $P>0$, that is we must have $$ \|\bm{x}_i\|^2 \le nP \qquad \forall i\in\{1,\ldots,K\}\,.$$

The quasi-static Rayleigh fading channel is defined similarly, except that each user's input is scaled by an independent channel gain $H_i$ , that is we have $\bm{Y} \in \mathbb{C}^n$ generated as $$ \qquad \bm{Y} = \sum_{i=1}^K H_i \bm{x}_i + \bm{Z}, \qquad H_i \stackrel{i.i.d.}{\sim} \mathcal{CN}(0,1)\,.$$ See~\cite[Sections 20.3 and 20.9]{polyanskiy2024information} for more on these channel models. Note that since we are primarily interested in the uncoordinated case, we cannot assume that $H_i$'s are known at the receiver.

\smallskip \textbf{Point-to-point channel codes.} A special case of $K=1$ is called a point-to-point channel (since there is only one transmitter and one receiver). An $(n,M,\epsilon)$ error-correcting code for a point-to-point channel is defined as a collection of $M$ codewords $\bm{c}_1,\ldots, \bm{c}_M$ together with a decoder function $g:\mathbb{C}^n \to [M] \triangleq \{1,\ldots,M\}$ such that the average probability of error satisfies $$ P_e = {1\over M} \sum_{m=1}^M \mathbb{P}[g(\bm{Y}) \neq m | \bm{X} = \bm{c}_m] \le \epsilon\,.$$ See~\cite[Chapter 17]{polyanskiy2024information} for more on the definition of the error-correcting codes.

It is known that for the \ac{AWGN} channel the best point-to-point codes satisfy~\cite{PPV10}: 
\begin{align*}
\log M &=  nC - \sqrt{nV} Q^{-1}(\epsilon) + O(\log n)\\ & \approx  nC - \sqrt{nV} Q^{-1}(\epsilon)
\end{align*}
where $Q^{-1}(\cdot)$ is the inverse of the $Q$-function, $C=\log (1+{P\over \sigma^2})$ is the channel capacity and $V={P (P+2\sigma^2)\over (P+\sigma^2)^2} \log^2 e$ is the channel dispersion.

Besides probability of error, another important figure is the normalized energy-per-bit, or $E_b\over N_0$, defined as $$ {E_b\over N_0} \triangleq {nP\over 2 \sigma^2 \log_2 M}$$ which quantifies the energy spent by a user terminal to transmit an information bit. See~\cite[Section 21.1]{polyanskiy2024information} for more on energy-per-bit.

\smallskip\textbf{UMAC codes.} Finally, we define the new type of error-correcting codes that will enable a principled exploration of random access with \ac{MPR}. An $(n, M, \epsilon, K_a)$ \ac{UMAC} code is a collection of $M$ codewords $\bm{c}_1,\ldots,\bm{c}_M \in \mathbb{C}^n$ and a decoder function $g:\mathbb{C}^n \to {[M]\choose K_a}$, where ${[M] \choose K_a}$ denotes a collection of all subsets of $K_a$ elements from the set $[M]$. The codebook and the decoder should satisfy the \textit{\ac{PUPE}} $\le \epsilon$ constraint. To define \ac{PUPE} we suppose that user $j$ selects uniformly at random (independently of other users) a message $W_j$ from $[M]$ and sets its channel input $\bm{x}_j = \bm{c}_{W_j}$. Once all $K_a$ channel codewords are selected they are input to a $K_a$-user channel (\ac{AWGN} or fading), which produces the output $\bm{Y}$. The decoder output $g(\bm{Y})$ is the list of messages that the decoder believes were transmitted by the users. The \ac{PUPE} is defined as\footnote{More exactly, to match the definition in~\cite{Polyanskiy2017}, we have to include in the error event also the case that $W_j$ clashes with a message $W_i$ for some $i\neq j$. However, since the chance of this happening is at most $K_a^2\over 2M$ and we are focused on $M=2^{100}$ in this survey, we prefer to omit this irrelevant term.} $$ \PUPE \triangleq {1\over K_a} \sum_{j=1}^{K_a} \PP[W_j \not \in g(\bm{Y})]\,. $$ That is, the \ac{PUPE} measures the probability that a user's message is going to be absent from the list of messages decoded by the receiver. See~\cite{Polyanskiy2017} for more formal details on \ac{UMAC} codes, as well as connections to related concepts in combinatorics and sparse regression.

Let us reflect on some of the ideas encoded in the mathematical definition above. First and foremost, despite having $K_a$ active users, there is only one codebook shared by all of them. This requirement formalizes the situation in which uplink transmission happens in an uncoordinated way (the only coordination is a common frame boundary signalled by the BS' beacon).

Second, while traditional error-correcting codes are required to arrange their codewords in a way that allows the identity of a transmitted point to be decodable from a noisy observation, the \ac{UMAC} code faces a more difficult challenge: a subset of any $K_a$ codewords should be decodable from observing a noisy sum of those codewords. In particular, since the users all employ the same codebook and the channel is invariant to permutation of the users, we can see that the decoder is not able to ever associate messages to users and is only recovering an unordered \textit{subset} of codewords. This is the reason for the name \textit{unsourced} MAC, since the messages are not sourced back to their originators. We remark, however, that as identity of the user is likely a part of the payload, this ambiguity is easy to resolve at higher layers. Leaving the problem unsourced, though, makes the coding-theoretic part cleaner and more natural.

Third, we observe an important departure from the classical \ac{MAC} in information theory: the \ac{PUPE} criterion only bounds the probability of error for an individual user. This choice is not only reasonable from the system-level point of view (since the uplink design criterion is to satisfy a certain accuracy for each separate user), but also natural theoretically. If an error is declared whenever any of the $K_a$ messages is misdecoded (as is done classically), the required $E_b/N_0$ grows without bound as $K_a$ increases, cf.~\cite[Slide 83]{polyanskiy2018information}.

In summary, the \ac{UMAC} code is defined in a way that formalizes the notion of uncoordinated random access by a-priori indistinguishable users and defines error probability in a way that allows analysis even for large $K_a$ without penalizing energy efficiency.

%%%%%%%%%% architectures %%%%%%%%%%%%%%%%%%%%%%%%%%%%%%%%%%%%%%%%%%%%%%%%%%%%%%%%%%%%%
%%%%%%%%%%%%%%%%%%%%%%%%%%%%%%%%%%%%%%%%%%%%%%%%%%%%%%%%%%%%%%%%%%%%%%%%%%%%%%%%%%%%%%
%%%%%%%%%%%%%%%%%%%%%%%%%%%%%%%%%%%%%%%%%%%%%%%%%%%%%%%%%%%%%%%%%%%%%%%%%%%%%%%%%%%%%%

\section{Prominent Unsourced Multiple Access Architectures}\label{sec:architectures}

Quickly after the introduction of the \ac{UMAC} setting, several coding schemes were proposed to approach the \ac{UMAC} performance limits, over the Gaussian \ac{MAC} as well as over the fading \ac{MAC}. Most of the constructions are directly inspired by the \ac{CS} perspective adopted in \cite{Polyanskiy2017}, and can be classified according to four emerging architectures, namely: Aloha-based schemes enhanced by some form of \ac{MPR} capability, \ac{CCS} schemes, preamble-based architectures, and spreading-based architectures. The different schemes are discussed in the following subsections. Given the rapidly evolving landscape of \ac{UMAC} code constructions, no attempt will be made to provide a thorough review of the existing methods. Rather, the focus will be on the distinctive features of the different architectures. For some selected schemes, results on the Gaussian \ac{MAC} with $n=30000$ channel uses are reported in Figure \ref{fig:all_res}.

\subsection{Multi-Packet Reception slotted Aloha Architectures}

Although the concept of \ac{MPR} in Aloha systems is relatively old (see Section \ref{subsec:FirstPeriod}), the design of coding mechanisms that allow the decoding of moderate-size collision sets has received renewed interest in the \ac{UMAC} context. The use of slotted Aloha as a basis for developing powerful \ac{UMAC} schemes stems from the following observation: Decoding collisions that involve many users according to the model of Section \ref{sec:UMAC} entails high complexity. On the contrary, moderate-size collision clusters can be resolved with low complexity using powerful error-correcting codes. In slotted Aloha, the average collision set size in a slot is a fraction of the number of active users---equal to the number of active users, divided by the number of slots in which the \ac{MAC} frame is partitioned. This enables the use of effective strategies to resolve multiple collisions in a slot.

The design of \ac{UMAC} schemes relying on this principle was introduced in \cite{OP:Low.2017}, where a $T$-fold Aloha construction was introduced. The construction implements a layered approach, where each message is encoded first by an outer code for $T$-users \ac{BAC}, and then by an inner binary linear code, designed to enable the decoding of the modulo-$2$ sum of the colliding codewords. The outer code can be based on the columns of the parity-check matrix of a $T$-error-correcting BCH code, while the inner code can be any powerful short-blocklength linear block code with low decoding complexity. The inner code decoder delivers with high probability the module-$2$ sum of the colliding codewords to the outer code decoder. If the cardinality of the collision set is within the decoding radius $T$ of the outer code, the set of collision messages is resolved. Otherwise, the collision batch is lost.

Effective \ac{MPR} mechanisms in  $T$-fold Aloha schemes for the Gaussian \ac{MAC} channel have been proposed in \cite{Narayanan2019:UMAC,Marshakov2019}, which rely on joint user decoding using \ac{LDPC} and polar codes, respectively. Both constructions also include packet repetition, allowing interference cancellation across slots in a way that is reminiscent of the \ac{IRSA} protocol \cite{Liva11:IRSA}. The performance of the scheme of \cite{Marshakov2019} is shown in Figure \ref{fig:all_res}. A $T$-fold Aloha construction was introduced in \cite{Duman21:trellis}, where the \ac{MPR} capability is provided by using short, terminated convolutional codes with joint decoding over a super-trellis tailored to the collision cluster. After convolutional encoding, randomized signature sequences are applied to the encoded packet. Their role is to enhance the performance of the multi-user trellis decoder.

\begin{figure*} 
	\centering 
	\subfloat[CCS: $\bm{x} = (\bm{x}_1,\bm{x}_2,\ldots,\bm{x}_L)$. \label{fig:CCS}]{\includegraphics[height=0.25\linewidth]{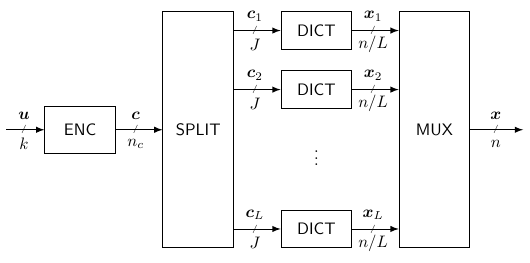}} \hspace{4mm} \subfloat[CCS, in the SPARC variant: $\bm{x} = \bm{x}_1 + \bm{x}_2 + \ldots \bm{x}_L$. \label{fig:SPARCS}]{\includegraphics[height=0.25\linewidth]{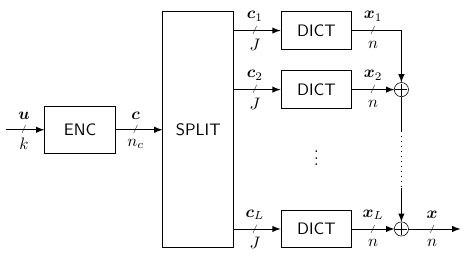}} \caption{Coded compressive sensing architectures.}\label{fig:CCSall} \end{figure*}

$T$-fold Aloha schemes also yield very promising results in fading channels. Joint user decoding and channel estimation on the quasi-static Rayleigh fading channel have been investigated in \cite{Frolov20}. The approach, which is based on \ac{LDPC} codes with multi-user \ac{BP} decoding, allows supporting moderate-size user populations with single-antenna receivers. A largely improved performance was achieved in \cite{Andreev2020} by replacing \ac{LDPC} codes with low rate polar codes, with a simple \ac{TIN}-\ac{SIC} decoder architecture. In fact, assuming independent fading coefficients, user separation turns out to be an easier task, allowing the use of low-complexity strategies based on single-user decoding and interference cancellation. A distinctive feature of the approaches of \cite{Frolov20,Andreev2020} is the lack of pilot sequences, with channel estimation performed directly on the transmitted data. A pilot-based $T$-fold Aloha scheme for massive multi-antenna systems was proposed in \cite{DBS+:Random.2017a}, where the adoption of randomized pilot hopping patterns, interleaved with the segments of the user codewords, was used to mitigate the pilot contamination effect that stems from the impossibility to allocate orthogonal pilot sequences to all users. In \cite{SDSP:Coded.2018} a related approach was introduced, employing sparse transmission patterns such as in the frameless Aloha \cite{Stefanovic12:Frameless,Stefanovic13:RatelessAloha} protocols. In \cite{Duman23,OKD:Slotted.2023}, a slotted Aloha scheme based on the construction of \cite{FMJC:PilotBased.2022} was proposed. Here, the pilot sequences are selected by the users from a common sensing matrix, and \ac{CS} techniques are used to obtain estimates of the channel coefficient vectors of the detected users. This class of schemes was shown to support a large number of active users with a \ac{SNR} that is only marginally larger than the one required in the single-user case.

\subsection{Coded Compressive Sensing Architectures}

Recognizing the elegance of the perspective outlined in \cite{Polyanskiy2017}, efforts to construct practical \ac{UMAC} schemes based on the \ac{CS} analogy were initiated in \cite{Narayanan2020,Calderbank2019} by introducing \ac{CCS} architectures.  \ac{CCS} uses a divide-and-conquer approach to address the exceedingly large dimensions of the sparse recovery problem associated with \ac{UMAC}. The principle underlying \ac{CCS} is to divide the message in $L$ small blocks of $J$ bits each, and to encode each block through a \ac{CS} dictionary of cardinality $2^J$, producing encoded blocks of length $n/L$. With $J$ in the order of a few bits, it is possible, at the receiver end, to apply low-complexity sparse recovery algorithms working on the $(n/L) \times 2^J$ sensing matrix obtained by stacking the $2^J$ dictionary sequences. Assuming a perfect recovery of the transferred blocks, the decoder is left with the task of ``stitching'' together the blocks associated with the individual user transmissions. The task can be accomplished by treating the outputs of the sparse recovery phase as observations of a so-called A-channel \cite{Chang1981} (see \cite{Fengler21} for an elegant casting of \ac{CCS} as a concatenated transmission scheme). Therefore, the code constructions for the A-channel can be used to reconstruct the individual user messages. For this purpose, in \cite{Narayanan2020} an ingenious construction of tree codes was proposed. A simplified description of the \ac{CCS} architecture as defined in \cite{Narayanan2020} is shown in Figure \ref{fig:CCS}.

\ac{CCS} in combination with \acp{SPARC} \cite{Joseph2012,Venkataramanan2019} was proposed in \cite{Fengler21}. Similarly to the approach of \cite{Narayanan2020}, each message is divided into $L$ blocks of $J$ bits each. Unlike \cite{Narayanan2020}, the $L$ blocks are encoded using a superposition code. According to the \ac{SPARC} construction, an $n \times L2^J$ matrix is partitioned into $L$ submatrices with dimensions $n\times 2^J$. The columns of each submatrix define a local dictionary that is used to encode a $J$-bits block. The encoded blocks are then added together. In \cite{Fengler21}, a modified \ac{AMP} decoder was introduced to extract the message blocks transmitted by the users, with the tree code of  \cite{Narayanan2020} used to reconstruct the user messages. The corresponding \ac{CCS} architecture is illustrated in Figure \ref{fig:SPARCS}. The performance on the Gaussian \ac{MAC} of the original \ac{CCS} scheme of \cite{Narayanan2020} (enhanced by a successive cancellation procedure, whereby decoded packets are subtracted from the signal at the input of the \ac{CS} recovery algorithm)  and the one of the \ac{SPARC}-based construction of \cite{Fengler21} are shown in Figure \ref{fig:all_res}. Remarkably, the latter shows a minimal increase of the required \ac{SNR} was the number of users grows, up to $150$ active users. It was soon recognized that the architecture of  Figure \ref{fig:SPARCS} is more general and it comprises the original architecture depicted in Figure \ref{fig:CCS} as a special case, where the sensing matrix is organized in a block diagonal form \cite{Ebert2021ICASSP}.

A \ac{SPARC}-based \ac{CCS} architecture with a modified outer code to allow joint decoding of the inner \ac{CS} code and of the outer code via iterative \ac{AMP}/BP was proposed in \cite{amalladinne2021unsourced}. The approach allows gains in the order of a few tenths of a dB over the scheme of \cite{Fengler21} in the moderate channel load regime.  Nonbinary \ac{LDPC} codes as outer codes with iterative \ac{BP}/\ac{AMP} decoding and denoising were proposed in \cite{EAR+:Coded.2022}. The use of an outer binary \ac{LDPC} code exploiting soft outputs delivered by a modified \ac{AMP} decoder has recently been explored in \cite{agostini2023bisparcs} in  a massive \ac{MIMO} setting, supporting a very large number of active users.

\subsection{Preamble-based Architectures}\label{sec:arch:preamble}

\begin{figure} 
	\centering    
	\includegraphics[width=0.95\columnwidth]{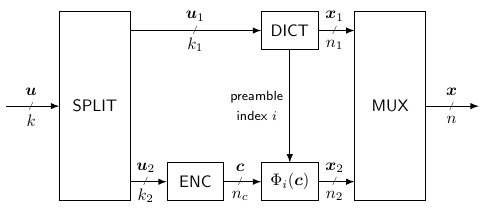} 
	\caption{Preamble-based: $\bm{x} = (\bm{x}_1,\bm{x}_2)$.} \label{fig:preamble-based} 
\end{figure}

Similarly to \ac{CCS} schemes, preamble-based  architectures implement an attack to the \ac{UMAC} problem by a divide-and-conquer approach. The idea is to apply a \ac{CS}-based detection of active users by means of preambles that users select from a moderate-size dictionary, thus, enabling low-complexity detection. The preambles are used to \emph{announce} the resources that will be used for data transmission in a subsequent phase. The principle is described in generic terms in Figure \ref{fig:preamble-based}. A first example of this class of \ac{UMAC} schemes was introduced in \cite{Narayanan:SIDMA}, where the preambles were associated with \ac{IDMA} access patterns \cite{IDMA}. The scheme, named sparse \ac{IDMA}, works as follows. User messages are divided into two parts. A first part, composed by $k_1$ bits, is used to address a dictionary of $2^{k_1}$ preambles. Denoting by $\bm{x}_1$ the selected preamble and by $i$ the corresponding index in the preamble dictionary, the index is used to select a repetition-and-interleaving pattern $\Phi_i$ (see Figure \ref{fig:preamble-based}). The remaining $k_2$ bits of the message are encoded by means of a $(n_c,k_2)$ binary linear block code, resulting in the codeword $\bm{c}$. The selected repetition-and-interleaving pattern $\Phi_i$ is applied to $\bm{c}$ generating a sparse vector $\bm{x}_2$ that is appended to the preamble $\bm{x}_1$. At the receiver side, the detection of transmitted preambles is performed through a suitable \ac{CS} algorithm. Given the list of detected preambles, an \ac{IDMA} decoding algorithm is employed to resolve the transmissions associated with the second part of the access frame.

The principle can be generalized to the use of strategies different from \ac{IDMA} for the second phase of transmission. In this sense, the mapping $\Phi_i$ in Figure \ref{fig:preamble-based} can be interpreted as a general transformation of the word $\bm{c}$.  In \cite{Truhachev21}, the mapping involves repetition, interleaving, and multiplication of the resulting vector by a binary signature, with the introduction of a randomized transmission delay. A similar construction, optimized for the block-fading Rayleigh channel, was introduced in \cite{NBT:Unsourced.2022}. Both in \cite{Narayanan:SIDMA} and in \cite{NBT:Unsourced.2022}, sophisticated multiuser decoding strategies are used to decode the second part of the detected transmissions. In \cite{Narayanan:SIDMA}, joint decoding is performed over the factor graph that couples the transmissions, according to the \ac{IDMA} patterns identified by the preambles. In \cite{NBT:Unsourced.2022}, iterative \ac{MUD} detection with soft interference cancellation is employed before decoding. The performance of the sparse \ac{IDMA} scheme of \cite{Narayanan:SIDMA}, where the $(n_c,k_2)$ binary linear block code is a binary \ac{LDPC} code optimized for the multiuser setting, is depicted in Figure \ref{fig:all_res}. The scheme works withing $2$ dB from the achievability bound up to $100$ active users.

A scheme closely related to sparse \ac{IDMA} was recently introduced in \cite{ODMA}, where the mapping $\Phi_i$ simply spreads the $n_c$ codeword bits over $n_c$ locations defined by the preamble index $i$. The scheme of \cite{ODMA} can be operated over the Gaussian \ac{MAC} without the transmission of the preamble, i.e., with a detection of the used access patterns that is performed through a likelihood ratio test that treats the codeword symbols as i.i.d. random variables in $\{-1,+1\}$. Nevertheless, to facilitate the detection of the access sequence and to enable the estimation of the channel on fading channels, the transmission of the selected preamble is considered in \cite{Ozates24}. Remarkably, the scheme of \cite{Ozates24} closely approaches the achievability bound of \cite{Polyanskiy2017} (see Section \ref{sec:UMAC}) on the Gaussian \ac{MAC} channel in the moderate load regime with a lean \ac{TIN}-\ac{SIC} receiver architecture, also thanks to the use of polar codes (concatenated with an outer CRC code).

We will see in Section \ref{sec:5GNR} that the grant-free mechanism recently introduced in the \ac{5GNR} standard adopts elements of the preamble-based architecture.

\subsection{Spreading-based Architectures}

The excellent match between random access and \ac{CDMA} and, more generally, spread spectrum techniques has been widely recognized, and it has been the basis of some of the most advanced random access protocols developed prior to the introduction of the \ac{UMAC} framework: the spread Aloha protocol \cite{Abramson90:VSAT}. Spread Aloha employs a unique spreading sequence for all users and relies on time asynchronism among users to facilitate detection and decoding. By tailoring the protocol to the \ac{UMAC} setting, it has been shown recently that spread Aloha with \ac{TIN}-\ac{SIC} decoding can closely approach the achievability bound on the Gaussian \ac{MAC} channel \cite{Schiavone24}.

Elegant spreading-based architectures explicitly designed for the \ac{UMAC} were introduced in \cite{Narayana20:polar,DLG:TensorBased.2021}. Regarding Aloha spread, both solutions are designed assuming perfect synchronization to the \ac{UMAC} framing structure, and rely on the use of multiple data-dependent spreading sequences.

\begin{figure} 
	\centering 
	\subfloat[Spreading-based: $\bm{x} = \bm{c} \otimes \bm{s}$. \label{fig:spreading-based}]{\includegraphics[height=0.30\linewidth]{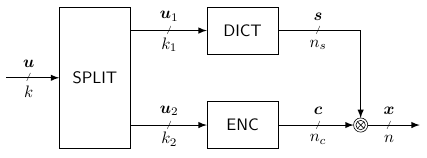}}\\ \vspace{4mm} 
	\subfloat[Spreading-based (tensor product): $\bm{x}=\bm{x}_L \otimes \bm{x}_{L-1}\otimes \dots \otimes \bm{x}_1$. \label{fig:spreading-based-tensor}]{\includegraphics[height=0.52\linewidth]{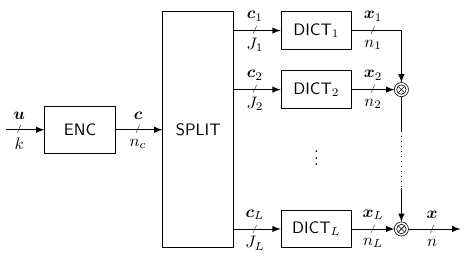}} \vspace{4mm} 
	\caption{Spreading-based architectures.}\label{fig:spreading-basedall} 
\end{figure}

\begin{figure*} 
	\centering 
	\includegraphics[width=1.55\columnwidth]{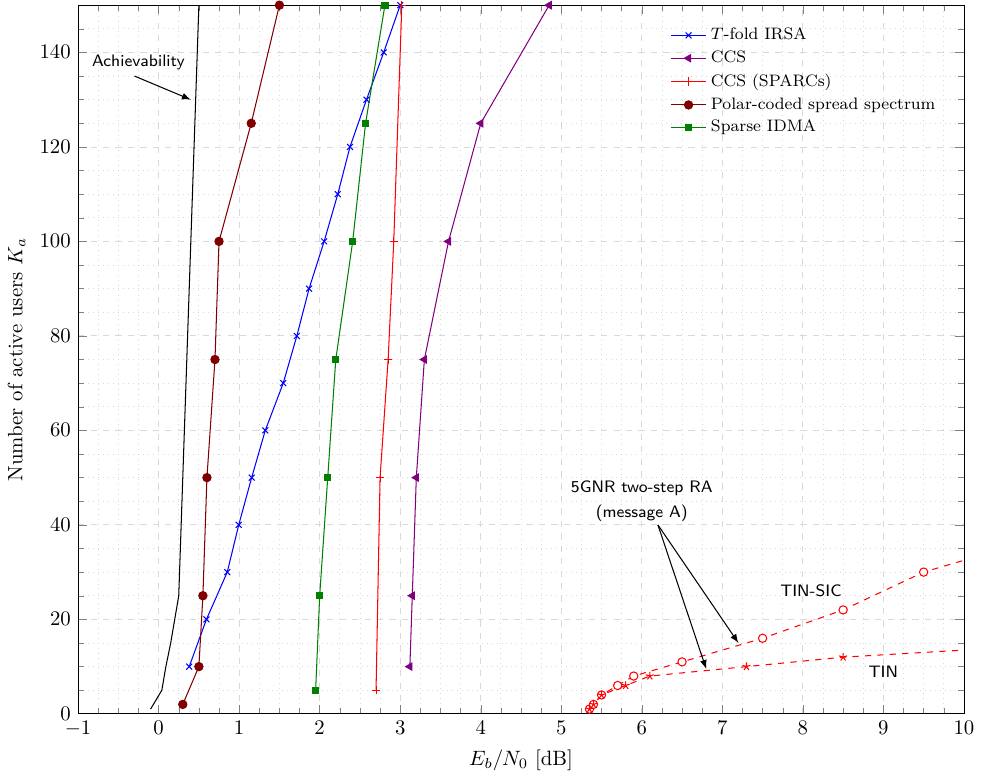} 
	\caption{Minimum \ac{SNR} required to achieve a $\PUPE = 5 \times 10^{-2}$, over the Gaussian \ac{MAC} channel. The frame length is $n=30000$ real channel uses. For the \ac{5GNR} two-step random access protocol, $n=32556$.}\label{fig:all_res} 
\end{figure*}

The construction of \cite{Narayana20:polar} is outlined in Figure \ref{fig:spreading-based}. Following the approach of preamble-based architectures, the user message is divided into two parts. A first part ($\bm{u}_1$, composed by $k_1$ bits, see Figure \ref{fig:spreading-based}) is used to select a spreading sequence from a dictionary. The second part of the user message is ($\bm{u}_2$, composed by $k_2$ bits) is encoded with a $(n_c,k_2)$ binary linear block code. Direct sequence spreading with the selected spreading sequence is applied to the codeword symbols. At the receiver side, energy detection is used to extract the set of spreading sequences that were selected by the active users. Decoding proceeds by linear \ac{MMSE} data estimation, followed by decoding and \ac{SIC}. The performance of the scheme on the Gaussian \ac{MAC}, with a polar code used to encode the second part of the message, is shown in Figure \ref{fig:all_res}. A remarkably small gap (less than $0.5$ dB) from the achievability bound can be observed, up to $100$ active users. Constructions similar to the one of \cite{Narayana20:polar} were introduced in \cite{HYX+:Sparse.2021a}, where sparse spreading codes were used, and in \cite{Gkagkos:2023}, where the original scheme \cite{Narayana20:polar} was engineered to operate on quasi-static fading channels with massive \ac{MIMO} arrays.

The spreading-based architecture introduced in \cite{DLG:TensorBased.2021} follows a less conventional path. In particular, a recursive encoding mechanism is used, with $L$ stages of spreading. The approach works as described in Figure \ref{fig:spreading-based-tensor}: the information message is first encoded by an outer code, resulting in the outer codeword $\bm{c}$, which is then divided in $L$ blocks $\bm{c}_1, \bm{c}_2, \ldots, \bm{c}_L$,  where the $i$th block comprises $J_i$ bits. Each block is encoded through a local dictionary (referred to as \emph{sub-constellation}), which is possibly different for each of the $L$ blocks. Denoting by $\bm{x}_i$ the local dictionary sequence that encodes $\bm{c}_i$, the output of the encoder is the tensor product of the sequences $\bm{x}_1, \bm{x}_2, \ldots, \bm{x}_L$. Joint decoding and channel estimation are performed on the decoder side, using rank-$1$ tensor decomposition to perform user separation. Single-user decoding follows for the detected users. The beauty of the architecture devised in \cite{DLG:TensorBased.2021} comes from its ability to separate users without relying on additional pilot sequences. The construction provides a competitive performance over block fading channels, with and without multi-antenna receivers.

\subsection{Maturity}

The four architectures described in the previous subsections depart from classical random access protocols that are found in many existing wireless communication systems. In this section we wanted to loosely rank the new architectures in terms of how ready they are for real-world deployment, in our opinion.

The vast majority of new architectures makes some use of \ac{CS} to solve part of the decoding problem: a few \ac{MPR} Aloha schemes employ \ac{CS} to detect pilot sequences of colliding users, \ac{CCS} builds on \ac{CS} to deliver an A channel to the outer tree code, preamble-based architectures use \ac{CS} techniques to identify preambles of active users, and certain spreading-based architectures use \ac{CS} to detect the spreading sequences selected by users. Most architectures foresee a layered approach to decoding. For example, in \ac{CCS} the receiver decodes using a serial approach that is typical of concatenated coding schemes: an inner decoder (based on \ac{CS}) outputs a set of sub-blocks associated to each slot, and an outer decoder has the task to stitch together the sub-blocks that form the message of each active user. In preamble-based architectures, the receiver first detects the preambles selected by the active users. This information is then used to direct a second decoding phase, which closely resembles the decoding of a coordinated non-orthogonal \ac{MAC} transmission scheme. It is certainly meaningful to refer to architectures that limit the differences to canonical random access protocols as more ``mature'', from an application viewpoint. Following this principle, \ac{MPR}-based Aloha architectures that rely on non-orthogonal pilot fields require very minor modifications to Aloha-based systems. Similarly, spreading-based architectures are natural candidates to improve the performance of spread Aloha systems. We will see in Section \ref{sec:5GNR} that preamble-based architectures are closely related to the random access protocols used by the \ac{5GNR} standard. These three architectures can be considered to be relatively mature. \ac{CCS}, on the contrary, relies on a more disruptive approach, which does not have any counterpart in adopted random access protocols. Therefore, the maturity of \ac{CCS} can be considered relatively low.

The maturity of the different approaches is also strongly related to the decoding / detection techniques required to harvest the performance gains. Schemes that rely mostly on single-user receiver chains --- possibly aided by interference cancellation --- exhibit a high level of maturity. In fact, \ac{SIC}-based receivers for random access protocols have been widely adopted in satellite communications \cite{etsi_smim,ETSI09b}. On the contrary, joint multiuser decoding (either in the form of joint \ac{LDPC} or polar decoding, or in the form of tree code decoding) is less explored in practical implementations.

Based on these considerations, we may provide a high-level discussion of the level of maturity for some of the schemes discussed in the previous subsections. Introducing a binary classification that focuses on the upcoming 3GPP standardization efforts --- ``6G-ready'' for schemes that present a high level of maturity, and ``beyond-6G'' for schemes that may require a longer engineering phase --- we summarize the discussion as follows.

\medskip

\textbf{MPR slotted Aloha architectures.} Schemes that rely on \ac{TIN}-\ac{SIC} decoders with slotted Aloha, using non-orthogonal pilot sequences for channel estimation \cite{DBS+:Random.2017a,SDSP:Coded.2018,Narayanan2019:UMAC,Duman23,OKD:Slotted.2023,FMJC:PilotBased.2022} can be considered 6G-ready. Schemes that rely on joint multiuser decoding \cite{Frolov20,Andreev2020} require further investigations in terms of hardware architecture, and hence can be categorized as beyond-6G.

\medskip

\textbf{CCS architectures.} \ac{CCS} architectures rely on a construction that is largely unexplored, in practical implementations. For this reason, we believe that \ac{CCS} represents a beyond-6G technology.

\medskip

\textbf{Preamble-based architectures.} Preamble-based architectures can be considered in general as quite mature. A distinction shall be made between schemes that require joint multiuser decoding \cite{Narayanan:SIDMA,NBT:Unsourced.2022}, and schemes that rely on single-user decoders, aided by \ac{SIC} \cite{Ozates24}. While the latter are 6G-ready, the former should be classified as beyond-6G. We will nevertheless see in Section \ref{sec:5GNR} that schemes closely resembling sparse \ac{IDMA} \cite{Narayanan:SIDMA} can be re-engineered to work with single user decoders (notably, largely reusing blocks already part of the \ac{5GNR} standard), retaining or even improving the remarkable performance of the original system, rendering sparse \ac{IDMA} a 6G-ready technology.

\medskip

\textbf{Spreading-based architectures.} The spreading-based construction of \cite{DLG:TensorBased.2021} performs joint decoding and channel estimation, exploiting rank-$1$ tensor decomposition to perform user separation. The receiver architecture is here highly innovative and sophisticated, and it is unexplored in system implementations. We classify the scheme as beyond-6G. On the contrary, upon retrieving the set of used spreading sequences, the construction of \cite{Narayana20:polar} reduces to a synchronous \ac{CDMA} scheme, whose implementation can be based on the numerous developments of \ac{CDMA} systems. We thus consider the spreading-based architecture of \cite{Narayana20:polar} as 6G-ready.

%%%%%%%%%% 6G evolutions %%%%%%%%%%%%%%%%%%%%%%%%%%%%%%%%%%%%%%%%%%%%%%%%%%%%%%%%%%%%%
%%%%%%%%%%%%%%%%%%%%%%%%%%%%%%%%%%%%%%%%%%%%%%%%%%%%%%%%%%%%%%%%%%%%%%%%%%%%%%%%%%%%%%
%%%%%%%%%%%%%%%%%%%%%%%%%%%%%%%%%%%%%%%%%%%%%%%%%%%%%%%%%%%%%%%%%%%%%%%%%%%%%%%%%%%%%%

\section{Grant-Free Access in Cellular Networks}\label{sec:5GNR}

\begin{figure} 
	\centering \subfloat[Four-step\label{fig:NR-a}]{\includegraphics[width=0.47\linewidth]{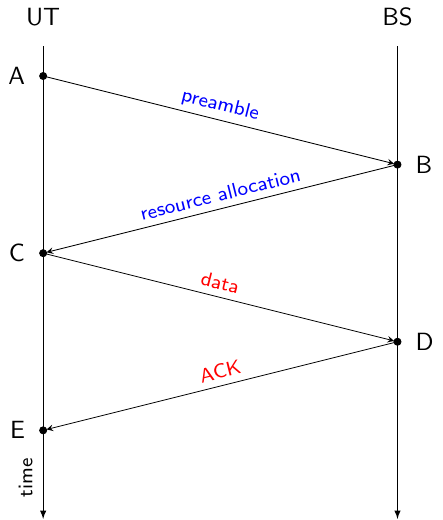}} \hfill \subfloat[Two-step\label{fig:NR-b}]{\includegraphics[width=0.47\linewidth]{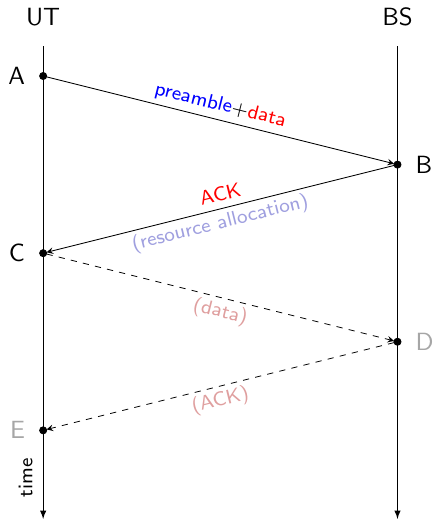}} \vspace{3mm} \caption{Random access procedures employed by \ac{LTE}/\ac{5GNR} standards. (a) Four-step random access: the user terminals (UT) transmit a preamble (\textsf{A}). Upon detection of the transmitted preambles, the base station (BS) provides a resource allocation to each detected user (\textsf{B}). UTs transmit their data packets in the allocated resources (\textsf{C}). The BS acknowledges correctly decoded packets (\textsf{D}). The procedure ends when the UT receives the acknowledgment (\textsf{E}). (b) Two-step random access (Release 16 of the \ac{5GNR} standard): A UT transmits a preamble, that directly points to the resource that will be used to transmit the data packet. The transmission of data packets follows without waiting for a resource allocation (\textsf{A}). At the BS, preambles are detected and decoding is attempted in the resources pointed by the preambles. For detected UT transmissions, an acknowledgment is sent to the UTs that are successfully decoded (\textsf{B}). For detected UT transmissions that do not result in successful decoding, orthogonal resources are allocated for the retransmission of the data packet, resuming the four-step random access procedure.}\label{fig:NR} 
\end{figure}

In the \ac{LTE} (including Narrowband IoT) \cite{LTE,NBIoT} and the \ac{5GNR} 3GPP standards \cite{5GNR15}, random access is based mainly on a four-step handshake between the user terminals and the \ac{BS}. The approach falls under the category of \emph{grant-based} random access protocols. Its behavior is illustrated in Figure \ref{fig:NR-a}, and it begins with the transmission of a preamble by the user terminal. In the \ac{5GNR} standard, the preamble is randomly chosen from a dictionary formed by $64$ Zadoff-Chu sequences, with a sequence length that can be set to $139$ or $839$ complex values (the preamble length and, possibly, preamble repetition policies are defined by the network configuration). Preamble transmission takes place in random access slots (\ac{PRACH}), where simultaneous preamble transmissions may collide. At the base station, preamble detection is performed, and feedback is sent to user terminals, which includes a resource allocation \,---\, in the form of a \ac{PUSCH} \emph{occasion} \,---\, for each detected user. The user terminals can then proceed with the transmission of their data packets in the assigned resource units. The base station finally acknowledges the correct reception of the packets.

\begin{figure*}[t] 
	\centering 
	\includegraphics[width=0.9\textwidth]{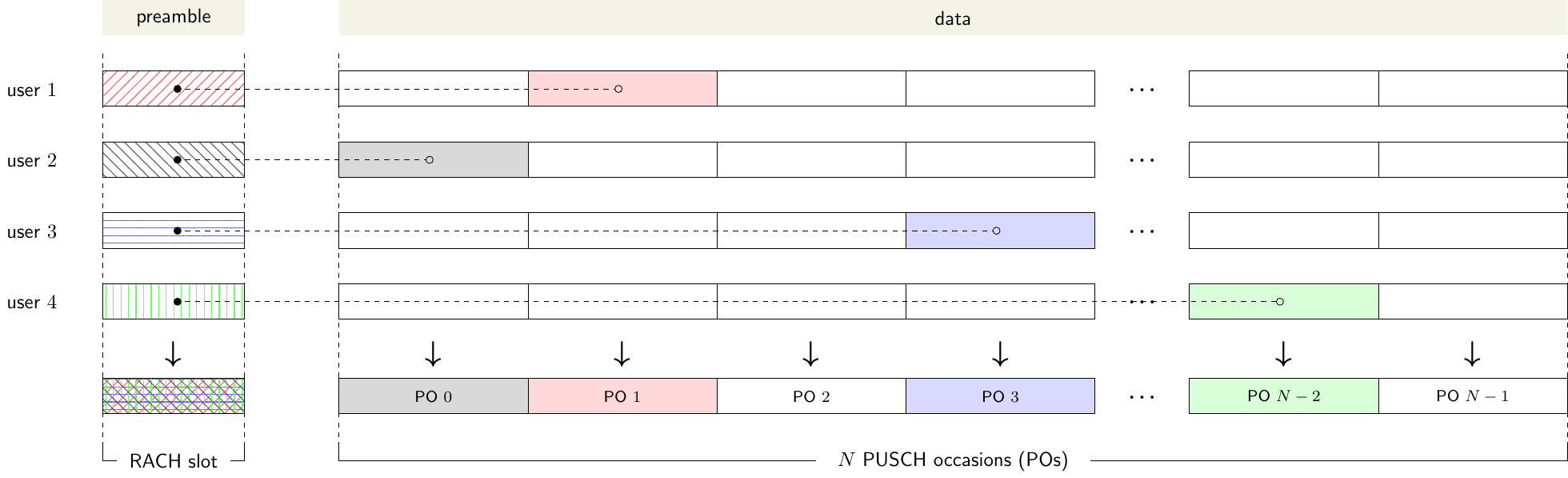} 
	\caption{Illustration of the two-step random access protocol of the \ac{5GNR} standard (first transmission only). The number of \ac{PUSCH} occasions is  $N$.}\label{fig:2step} 
\end{figure*}

With Release 16 of the \ac{5GNR} standard, a new random access procedure has been included, which incorporates elements of \emph{grant-free} access schemes \cite{5GNR16}. The protocol, commonly known as \emph{two-step random access}, was introduced with the main objective of reducing the delay entailed by the legacy four-step approach. The \ac{5GNR} two-step random access protocol works according to the preamble-based architecture described in \ref{sec:arch:preamble}. More specifically, a set of $N$ resources (\ac{PUSCH} occasions) is associated with the $64$ preambles from the same dictionaries used by the four-step procedure. Both a one-to-one mapping (with preambles pointing distinct resource units) and many-to-one mappings (with multiple preambles pointing to the same resource unit) are possible. User terminals initiate their transmission by selecting a preamble to be transmitted within a random access slot. Each preamble ``announces'' the \ac{PUSCH} occasion that will be used for data transmission by the user terminal, as detailed in Figure \ref{fig:2step}. The first step (referred to as \emph{message A} transmission in the standard) is completed by transmission of the packet in the selected resource unit. The base station performs preamble detection and attempts decoding of the packets transmitted within the \ac{PUSCH} occasions that were signaled by the detected preambles. If decoding succeeds, the two-step procedure ends with the base station acknowledging the correct reception to the user terminal (\emph{message B} transmission in the \ac{5GNR} jargon). If decoding fails, \ac{5GNR} two-step random access protocol allows to resume the four-step procedure, i.e., a negative acknowledgment if sent to the user terminal, contextually with a resource allocation.

Both the four- and the two-step random access protocols envisaged by the \ac{5GNR} standard are clearly not designed to support massive IoT networks. This observation stems from the very limited number of preambles,  which is almost two orders of magnitude smaller than the cardinality of the dictionaries adopted by more advanced schemes based on the same preamble-based architecture (see Section \ref{sec:arch:preamble}).

With the perspective of equipping future releases of the 3GPP standard with a fully-fledged grant-free \ac{MAC} protocol, and recognizing that standard updates favor solutions that have a limited impact on the system architecture, a sensible question is whether the message A transmission phase of the two-step random access protocol---or modifications of it---can be used to support massive connectivity scenarios. This critical question has recently been addressed in \cite{whitepaper2024}. The next subsection reviews some of the conclusions of that study.

\subsection{Two-Step Random Access: Performance}

Figure \ref{fig:all_res} shows the performance of two-step random access (message A transmission) over the Gaussian \ac{MAC}. The analysis is given for a specific configuration of the scheme: The short Zadoff-Chu preamble family is used, with $64$ sequences of length $139$ (complex channel uses), repeated twice (according to the A1 short preamble configuration of the \ac{5GNR} standard, see \cite[Chapter 16]{DAHLMAN2018}). The data packet encodes $100$ information bits in $500$ codeword bits, which are then mapped onto\ac{QPSK} symbols, resulting in $250$ channel uses. The $(500,100)$ \ac{5GNR} \ac{LDPC} code (from base matrix two) is used for encoding \cite{Ric18}. Taking into account $N=64$ available \ac{PUSCH} occasions, the frame length is $n = 2\times 139 + 64 \times 250 = 16278$ (equivalent to $32556$ real degrees of freedom). At the receiver side, the \ac{OMP} algorithm \cite{PatiOMP,TroppOMP} is used to detect preamble transmissions. The decoding of a packet is attempted at the \ac{PUSCH} occasions pointed by the detected preambles. The performance is evaluated with and without the aid of \ac{SIC}, which, when present, is again assumed to be ideal. The case where no \ac{SIC} is applied is referred to as \ac{TIN}, whereas \ac{TIN}-\ac{SIC} referrers to the application of interference cancellation. When \ac{SIC} is employed, preamble detection is repeated after canceling the interference contribution caused by the preambles associated with successful decoding attempts.

At low channel loads, the minimum \ac{SNR} required to achieve the target $\PUPE = 5\times 10^{-2}$ is largely influenced by the energy overhead introduced by the preamble (approximately $3.2$ dB), and by the inherent suboptimality of the  $(500,100)$ \ac{5GNR} \ac{LDPC} code, which exhibits a loss of $\approx 1.6$ dB with respect to single-user finite blocklength achievability bounds. Note that the first phase of the protocol behaves as slotted Aloha, with the addition of the preamble to announce \ac{PUSCH} occasions used for transmission. A classical slotted Aloha receiver would proceed by trying to decode each of $N$ resource units \,---\, possibly, by preceding the actual decoding attempts by an energy detection test, to discard \ac{PUSCH} occasions that do not contain transmissions. From this point of view, the energy spent in the preamble could be saved, greatly improving the efficiency of the protocol. However, the possibility of triggering the four-step collision resolution procedure requires the identification of user transmissions, which is made possible by the use of the preambles, which justified their use in the two-step random access protocol \cite{TWC+:Design.2021}. As the channel load increases, the minimum \ac{SNR} required to attain the target \ac{PUPE} grows quickly. When plain \ac{TIN} decoding is applied, the system struggles to support more than $K_a=8$ active users. Under \ac{TIN}-\ac{SIC} decoding, the situation improves. However, at moderate-large numbers of active users (e.g., $K_a = 30$) the scheme operates at more than $8$ dB from the achievability bound. The result can be explained by the limited ability of the \ac{LDPC} code to decode correct messages in the presence of multiple collisions, and by the low number of available \ac{PUSCH} occasions (at most $64$, according to the maximum number of preambles).

\begin{figure*}
	\centering
	\includegraphics[width=1.55\columnwidth]{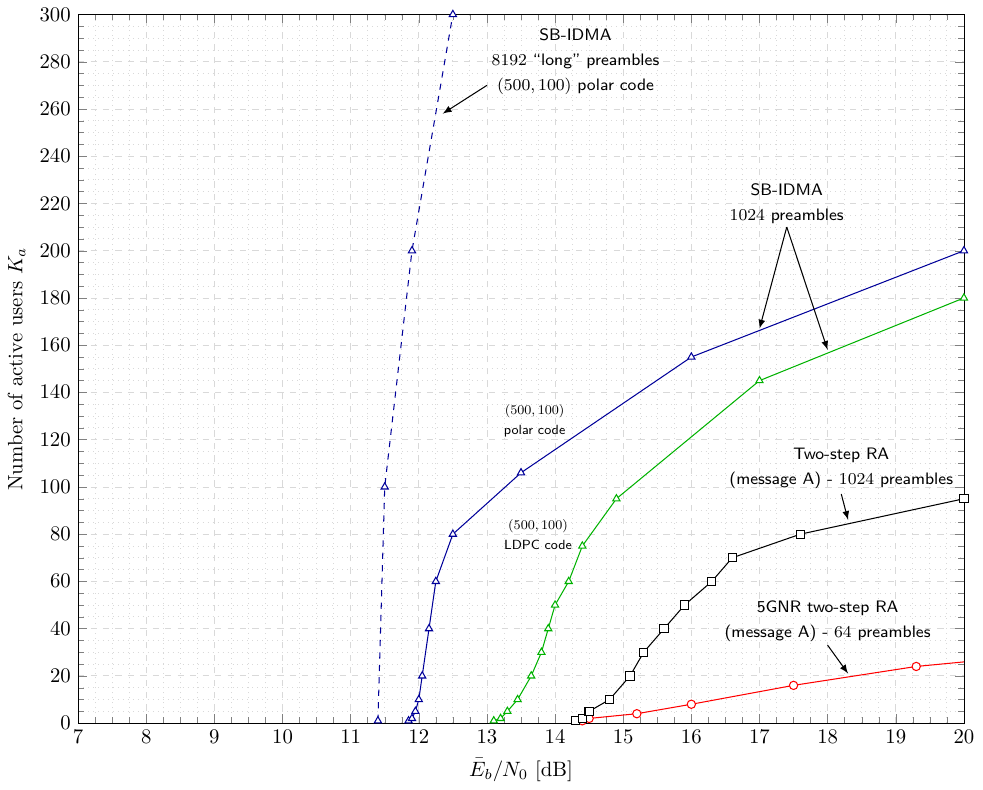}	\caption{Minimum average \ac{SNR} required to achieve a $\PUPE = 10^{-1}$, over the quasi-static Rayleigh fading \ac{MAC} channel. The frame length is $n=19478$ complex channel uses.} \label{fig:all_res_SISO_fading}
\end{figure*}

Although the results on Gaussian \ac{MAC} provide a first understanding of the limitations of the two-step random access protocol, the real test bench is transmission over fading channels. Figure \ref{fig:all_res_SISO_fading} reports the performance on the quasi-static Rayleigh channel. The configuration used over the Gaussian \ac{MAC} is preserved, with the only addition of a pilot field appended to each packet, which is used at the receiver end to estimate the fading channel coefficient (assumed to be independent between users). Therefore, the \ac{PUSCH} occasions are modified to accommodate $300$ channel uses. The overall frame length is $n = 2\times 139 + 64 \times 300 = 19478 $. The target \ac{PUPE} is set to $10^{-1}$. The size of the pilot field amounts to $50$ channel uses. The performance (provided only under \ac{TIN}-\ac{SIC}) shows again that two-step random access can hardly support large channel loads.

\subsection{Evolution Towards Full Grant-Free Random Access}

A simple modification of the scheme, obtained by enlarging the preamble set to $1024$ sequences, allows for a remarkable improvement in the supported channel load. Note that the result is obtained while the frame structure remains unchanged, with $64$ \ac{PUSCH} occasions of channel $300$ uses each. The reason for the improvement lies in the more accurate channel estimation indirectly provided by the enlarged preamble family. Each preamble points to a \ac{PUSCH} occasion \emph{and} to a specific pilot sequence: When two users pick different preambles that point to the same \ac{PUSCH} occasion, by employing two distinct pilot sequences, it is possible to extract sufficiently accurate estimates of the two channel coefficients. This allows to decode with high probability colliding transmissions. A small number of preambles (as in the standard two-step random access scheme) hinders the possibility of having a sufficient diversity of preamble sequences, thus reducing the \ac{MPR} capability of the receiver. Interestingly, in \cite{whitepaper2024} it was observed that very limited gains could be appreciated by increasing the size of the preamble family beyond $1024$. The result could be explained by an additional shortcoming of two-step random access, that is, the limited amount of access patterns ($64$) allowed by the construction. Inspired by the sparse \ac{IDMA} scheme of \cite{Narayanan:SIDMA}, a modification of the two-step random access protocol that relies on packet repetition is outlined in \cite{whitepaper2024}. With each packet repeated $\rho$ times and assuming again $N=64$ \ac{PUSCH} occasions, up to  $\binom{N}{\rho}$ access patterns can be defined. This allows to make a better use of a larger preamble sets. The performance of this scheme, referred to as \ac{SB-IDMA}, is provided in \ref{fig:all_res_SISO_fading}. The results are reported assuming either the \ac{5GNR} $(500,100)$ \ac{LDPC} code \cite{Ric18} or the \ac{5GNR} $(500,100)$ polar code \cite{Bio20} (the dimension of the code refers to the dimension of the outer $11$-bits CRC code). The  polar code is decoded via the \ac{SCL} algorithm \cite{TV15}, with a list size set to $128$. Large gains can be observed with respect to two-step random access, with a saturation of the performance at high \ac{SNR} that is mainly driven by the preamble misdetection probability. Additional gains can be noticed by tuning the \ac{SB-IDMA} scheme parameters: The result is achieved by  increasing the length of the preamble to $1778$ channel uses, enlarging the preamble set to $8192$ preambles, while simultaneously reducing the transmission power of the preamble by a factor $1/12$ and the number of \ac{PUSCH} occasions to $59$ (keeping the overall frame size fixed to $19478$). With this dimensioning of the scheme, large channel loads can be supported at a \ac{SNR} that is only a few tenths of a dB larger than the one required by single-user transmission. The results also highlight the performance improvement that can be achieved by adopting polar codes to protect the data packets. This fact stems from the superior performance of polar codes (with an outer CRC code and under \ac{SCL} decoding) with respect to \ac{LDPC} codes, in the short blocklength regime. We expect that the advantage in using polar codes may diminish when transmitting larger packets.

The analysis pinpoints two main limitations of the message A transmission phase of the \ac{5GNR} two-step random access protocol, namely (i) the small number of available preamble sequences and (ii) the limited number of access patterns that can be realized. Therefore, it is conceivable that evolutions of 3GPP standards aiming at massive random access scenarios should target the introduction of larger preamble families, as well as richer configurations of access patterns. A fundamental question is how the preamble family should be designed, especially considering the important role that the structure of Zadoff-Chu sequences plays in handling the different propagation delays of user terminals \cite[Chapter 11]{DAHLMAN2016285}.

%%%%%%%%%% hystory %%%%%%%%%%%%%%%%%%%%%%%%%%%%%%%%%%%%%%%%%%%%%%%%%%%%%%%%%%%%%%%%%%%
%%%%%%%%%%%%%%%%%%%%%%%%%%%%%%%%%%%%%%%%%%%%%%%%%%%%%%%%%%%%%%%%%%%%%%%%%%%%%%%%%%%%%%
%%%%%%%%%%%%%%%%%%%%%%%%%%%%%%%%%%%%%%%%%%%%%%%%%%%%%%%%%%%%%%%%%%%%%%%%%%%%%%%%%%%%%%

\section{Conclusions}\label{sec:conc}

The recent introduction of the unsourced multiple access (UMAC) perspective has originated a wave of new massive random access protocols, revolutionizing the landscape of large-scale uncoordinated multiple access communications. While the ripples of this wave settle, the crucial question of which UMAC coding architecture will emerge as the reference in future cellular wireless network standards remains open. Scalability, flexibility, and robustness with respect to a broad range of channel conditions and user terminal limitations will play a major role in defining the solution, calling for strict cooperation of coding and information theorists, wireless communication engineers, and hardware architecture designers, in a coordinated effort to refine state-of-the-art UMAC protocol designs.

%%%%%%%%%%%%%%%%%%%%%%%%%%%%%%%%%%%%%%%%%%%%%%%%%%%%%%%%%%%%%%%%%%%%%%%%%%%%%%%%%%%%%%
%%%%%%%%%%%%%%%%%%%%%%%%%%%%%%%%%%%%%%%%%%%%%%%%%%%%%%%%%%%%%%%%%%%%%%%%%%%%%%%%%%%%%%

\section*{Acknowledgments}
The authors would like to thank Kirill Andreev, Jean-Francois Chamberland, Alexander Fengler, Alexey Frolov, and Khrishna Narayanan for providing numerical results. The authors would also like to thank Maxime Guillaud for the insightful discussions on the tensor-based modulation of \cite{DLG:TensorBased.2021} and Bobak Nazer for detailed reading of the earlier draft.

%%%%%%%%%% bibliography %%%%%%%%%%%%%%%%%%%%%%%%%%%%%%%%%%%%%%%%%%%%%%%%%%%%%%%%%%%%%%
%%%%%%%%%%%%%%%%%%%%%%%%%%%%%%%%%%%%%%%%%%%%%%%%%%%%%%%%%%%%%%%%%%%%%%%%%%%%%%%%%%%%%%
%%%%%%%%%%%%%%%%%%%%%%%%%%%%%%%%%%%%%%%%%%%%%%%%%%%%%%%%%%%%%%%%%%%%%%%%%%%%%%%%%%%%%%
%%%%%%%%%%%%%%%%%%%%%%%%%%%%%%%%%%%%%%%%%%%%%%%%%%%%%%%%%%%%%%%%%%%%%%%%%%%%%%%%%%%%%%
%%%%%%%%%%%%%%%%%%%%%%%%%%%%%%%%%%%%%%%%%%%%%%%%%%%%%%%%%%%%%%%%%%%%%%%%%%%%%%%%%%%%%%

\balance 

% Generated by IEEEtran.bst, version: 1.13 (2008/09/30)

%%%%%%%%%%%%%%%%%%%%%%%%%%%%%%%%%%%%%%%%%%%%%%%%%%%%%%%%%%%%%%%%%%%%%%%%%%%%%%%%%%%%%%
%%%%%%%%%%%%%%%%%%%%%%%%%%%%%%%%%%%%%%%%%%%%%%%%%%%%%%%%%%%%%%%%%%%%%%%%%%%%%%%%%%%%%%

\end{document}